\documentclass[prb,aps,twocolumn,showpacs,nofootinbib]{revtex4}
\usepackage{graphicx} 
\usepackage{dcolumn}  
\usepackage{bm}       
\usepackage{amsmath}
\usepackage{epsfig}

\def\ii{{\rm i}}      
  \def\Rb{{\bf R}}  \def\pb{{\bf p}}  
    \def\Eb{{\bf E}}  \def\rb{{\bf r}}  \def\db{{\bf d}}
\def\xx{\hat{\bf x}}     
\def\yy{\hat{\bf y}}     
\def\zz{\hat{\bf z}}   \def\ph{\hat{p}}   \def\qh{\hat{q}}   \def\Hh{\hat{H}}   \def\ah{\hat{a}}
\def\eh{\hat{\bf{\varepsilon}}}

\begin{document}
\title{Vacuum friction in rotating particles}
\author{A.~Manjavacas and F.~J.~Garc\'{\i}a~de~Abajo}
\email[Corresponding author: ]{J.G.deAbajo@csic.es}
\affiliation{Instituto de \'Optica - CSIC, Serrano 121, 28006 Madrid, Spain}

\date{\today}

\begin{abstract}
We study the frictional torque acting on particles rotating in empty space. At zero temperature, vacuum friction transforms mechanical energy into light emission and produces particle heating. However, particle cooling relative to the environment occurs at finite temperatures and low rotation velocities. Radiation emission is boosted and its spectrum significantly departed from a hot-body emission profile as the velocity increases. Stopping times ranging from hours to billions of years are predicted for materials, particle sizes, and temperatures accessible to experiment. Implications for the behavior of cosmic dust are discussed.
\end{abstract}
\maketitle

\section{Introduction}

The radiation emitted by accelerated charges produces reaction forces acting back on them. For rotating charged particles (e.g., electric\cite{B1988} and magnetic\cite{P1955} dipoles), this gives rise to reaction torques.\cite{C1967} Likewise, accelerated neutral bodies are known to experience friction because they emit light due to the absolute change in the boundary conditions of the electromagnetic field. This is the so-called Casimir radiation.\cite{KG99,KN02}

A spinning sphere presents a more challenging situation: its surface appears to be unchanged, although it experiences a centripetal acceleration. So, the question arises, does a homogeneous, neutral sphere emit light simply by rotating? Is such a particle slowing down when spinning in vacuum? We know the inverse process to be true: the angular momentum carried by light can be transformed into mechanical rotation of neutral particles.\cite{TYL05} However, this type of problem requires a delicate analysis, somehow related to the non-contact friction predicted to occur between planar homogeneous surfaces set in relative uniform motion,\cite{P97} which is currently generating a heated debate.\cite{controversy}

In this paper, we investigate the friction produced on rotating neutral particles by interaction with the vacuum electromagnetic fields. Friction is negligible in dielectric particles possessing large optical gap compared to the rotation and thermal-radiation frequencies. For other materials (e.g., metals), in contrast to previous predictions,\cite{P06} we find nonzero stopping even at zero temperature. The dissipated energy is transformed into radiation emission and thermal heating of the particle, although cooling relative to the surrounding vacuum is shown to take place under very common conditions. We formulate a theory that describes these phenomena and allows us to predict experimentally measurable effects.

\section{Theoretical description}

\begin{figure}
\centerline{\includegraphics*[width=7cm]{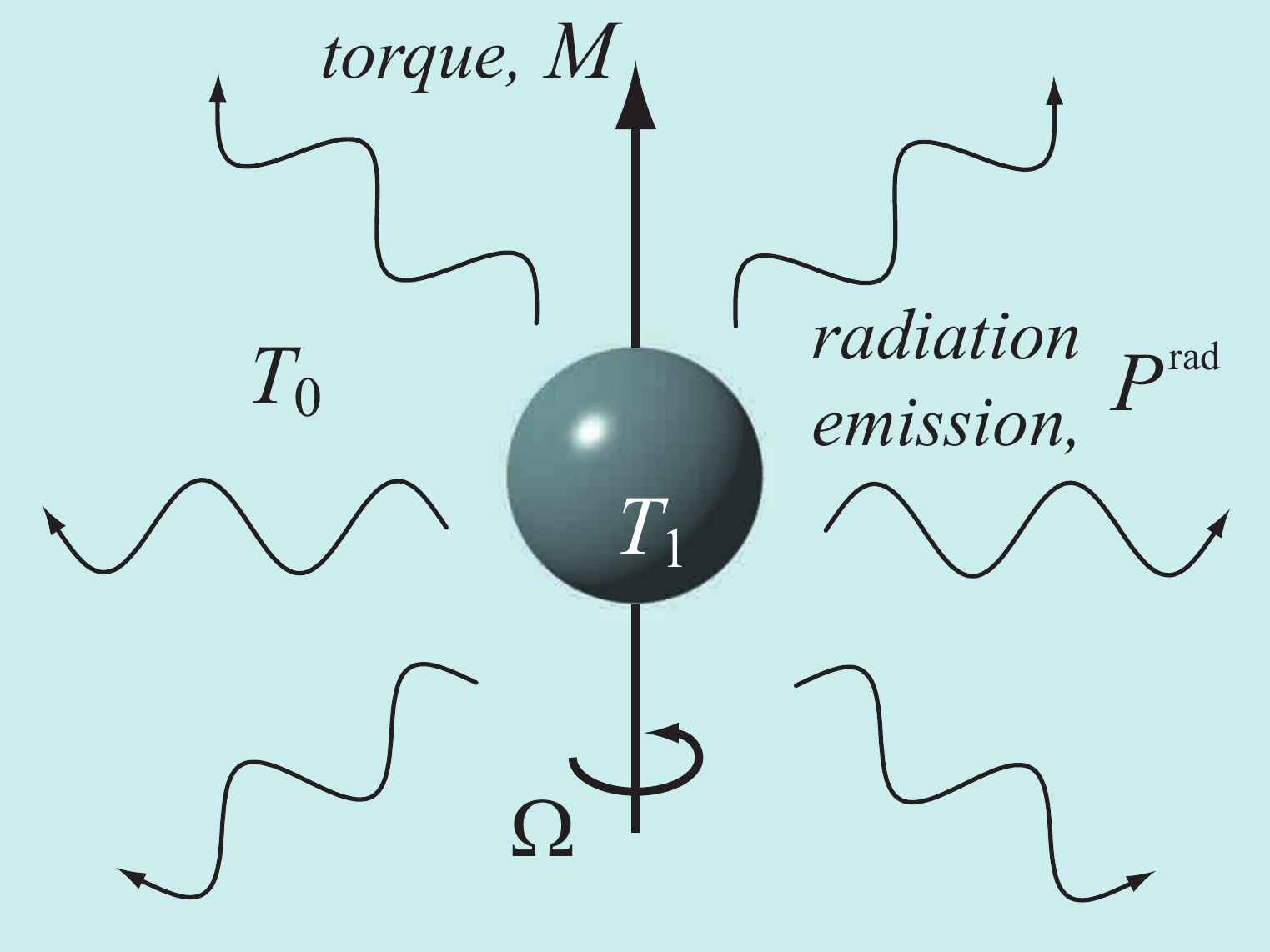}}
\caption{Sketch of a spherical rotating particle and parameters considered in this work. The particle is at temperature $T_1$ and rotates with frequency $\Omega$. The interaction with vacuum at temperature $T_0$ produces a frictional torque $M$ and a radiated power $P^{\rm rad}$.} \label{Fig1}
\end{figure}

We consider an isotropic particle at temperature $T_1$ spinning with frequency $\Omega$ and embedded in a vacuum at temperature $T_0$ (see Fig.\ \ref{Fig1}). The particle experiences a torque $M$ by interaction with the surrounding radiation field and it is also capable of exchanging photons, with net emission power $P^{\rm rad}$. For simplicity, we assume the particle radius $a$ to be small compared to the wavelength of the involved photons, so that we can describe it through its frequency-dependent polarizability $\alpha(\omega)$. Since the maximum frequency of exchanged photons is controlled by the rotation frequency and the thermal baths at temperatures $T_0$ and $T_1$, this approximation implies that both $\Omega a/c$ and $k_BT_ja/c\hbar$ are taken to be small compared to unity. These conditions are fulfilled in very common situations (for instance, for $a=50\,$nm, one has  $\Omega\ll6\times10^3\,$THz and $T_j\ll4.6\times10^4\,$K).

Friction originates in fluctuations of both (i) the vacuum electromagnetic field $\Eb^{\rm fl}$ and (ii) the particle polarization $\pb^{\rm fl}$. We calculate the emitted power from the work exerted by the particle dipole,
\begin{eqnarray}
P^{\rm rad}=-\left\langle\Eb^{\rm ind}\cdot\partial\pb^{\rm fl}/\partial t+\Eb^{\rm fl}\cdot\partial\pb^{\rm ind}/\partial t\right\rangle,\label{P00}
\end{eqnarray}
where $\Eb^{\rm ind}$ is the field induced by $\pb^{\rm fl}$, and $\pb^{\rm ind}$ is the dipole induced by $\Eb^{\rm fl}$. Likewise, the torque is obtained from the action of the field on the dipole,
\begin{eqnarray}
{\bf M}=\left\langle\pb^{\rm fl}\times\Eb^{\rm ind}+\pb^{\rm ind}\times\Eb^{\rm fl}\right\rangle.\label{M00}
\end{eqnarray}
The result is quadratic in $\Eb^{\rm fl}$ for contribution (i) and in $\pb^{\rm fl}$ for contribution (ii). The brackets $\left\langle\right\rangle$ represent the average over these quadratic fluctuation terms, which we perform using the fluctuation-dissipation theorem (FDT) (see Appendix).

Rotational motion enters here through the transformation of the field and the polarization back and forth between rotating and lab frames. This is needed because the particle polarizability can only be applied in the rotating frame, in which the electronic and vibrational excitations participating in $\alpha$ are well defined and $\Omega$-independent. In contrast, the effective polarizability in the lab frame has a dependence on $\Omega$. Further details of this formalism are given in the Appendix. The resulting radiated power reads (see Appendix for a detailed derivation)
\begin{eqnarray}
P^{\rm rad}=\int_{-\infty}^\infty\hbar\omega\;d\omega\;\Gamma(\omega)\label{P},
\end{eqnarray}
where
\begin{eqnarray}
\Gamma(\omega)&=&(2\pi\omega\rho^0/3)\,\Big\{2g_\perp(\omega-\Omega)\,\big[n_1(\omega-\Omega)-n_0(\omega)\big]\nonumber\\
&&+g_\parallel(\omega)\,\big[n_1(\omega)-n_0(\omega)\big]\Big\}\label{Pw}
\end{eqnarray}
is the spectral distribution of the rate of emission (when $\omega\Gamma>0$) or absorption ($\omega\Gamma<0$),
$n_j(\omega)=[{\exp(\hbar\omega/k_BT_j)-1}]^{-1}$ is the Bose-Einstein distribution function at temperature $T_j$,
\begin{eqnarray}
g_l(\omega)={\rm Im}\{\alpha_l(\omega)\}-\frac{2\omega^3}{3c^3}|\alpha_l(\omega)|^2\nonumber
\end{eqnarray}
are odd functions of $\omega$ describing particle absorption for polarization either parallel ($l=\parallel$) or perpendicular ($l=\perp$) with respect to the rotation axis, and $\rho^0=\omega^2/\pi^2c^3$ is the free-space local density of photonic states. These results apply to particles with orthogonal principal axes of polarization, rotating around one of them, and with $\alpha_\perp$ given by the average of the polarizability over the remaining two orthogonal axes. The torque $M$ takes a similar form,
\begin{eqnarray}
M=-\int_{-\infty}^\infty d\omega\;\hbar\Gamma(\omega).\label{M}
\end{eqnarray}
Incidentally, the $g_\parallel$ term of Eq.\ (\ref{Pw}) vanishes under the integral of Eq.\ (\ref{M}), and furthermore, $M=0$ for $\Omega=0$. In the $T_0=T_1=0$ limit, one has $n_j(\omega)=-\theta(\omega)$, from which we find the integrals to be restricted to the $(0,\Omega)$ range: only photons of frequency below $\Omega$ can be generated.

Unfortunately, Eqs.\ (\ref{P00}) and (\ref{M00}) do not account for radiative corrections coming from the elaborate motion of induced charges in the rotating particle. Although such corrections are insignificant for small particles, we incorporate them here for spheres in a phenomenological way through the term proportional to $|\alpha|^2$ in Eq.\ (\ref{Pw}), preceded by a coefficient chosen to yield $g_l=0$ (and consequently, $M=0$) in non-absorbing particles:\cite{V1981} internal excitations (i.e., absorption) are necessary to mediate the coupling between the rotational state and radiation.\cite{friction2} Furthermore, we neglect magnetic polarization, which can be important for large, highly conductive particles.\cite{frictionn}

\section{Metallic particles}

This case is representative for absorbing particles. At low photon frequencies $\omega$ below the interband transitions region, metals can be well described by the Drude model, characterized by a DC electric conductivity $\sigma_0$ and a dielectric function $\epsilon=1+\ii\,4\pi\sigma_0/\omega$.\cite{AM1976} For a spherical particle of radius $a$, we have $\alpha\approx a^3(\epsilon-1)/(\epsilon+2)$, and consequently
\begin{eqnarray}
{\rm Im}\{\alpha(\omega)\}\approx3\omega a^3/4\pi\sigma_0.\label{drude}
\end{eqnarray}
For sufficiently small particles, absorption dominates over radiative corrections, so that we can overlook terms proportional to $|\alpha|^2$ in Eqs.\ (\ref{P})-(\ref{M}). Then, we find the closed-form expressions
\begin{eqnarray}
P^{\rm rad}_{\rm D}=\frac{\hbar a^3}{60\pi^2c^3\sigma_0}\Big[2\Omega^6+5\Omega^4\theta_1^2+3\Omega^2\theta_1^4
+\frac{5}{14}(\theta_1^6-\theta_0^6)\Big]
\label{Pdrude}
\end{eqnarray}
and
\begin{eqnarray}
M_{\rm D}=\frac{-\hbar a^3\Omega}{120\pi^2c^3\sigma_0}\Big[6\Omega^4+10\Omega^2\theta_1^2+\theta_0^4+3\theta_1^4\Big],
\label{Mdrude}
\end{eqnarray}
where the subscript D refers to the Drude model and
\begin{eqnarray}
\theta_j=2\pi k_BT_j/\hbar.\nonumber
\end{eqnarray}
Equations\ (\ref{Pdrude}) and (\ref{Mdrude}) show that vacuum friction is always producing stopping ($M\Omega<0$), whereas the balance of radiation exchange between particle and free space can change sign depending on their relative temperatures. The general trend of these expressions is shown in Fig.\ \ref{Fig2}(b). At low $\Omega$, the torque scales as $\Omega$, whereas a steeper $\Omega^5$ dependence is observed at faster velocities. Interestingly, a nonzero torque $M\propto\Omega^5$ is predicted at $T_0=T_1=0$, despite the axial symmetry of the particle.

\section{Equilibrium temperature}

\begin{figure}
\centerline{\includegraphics*[width=8cm]{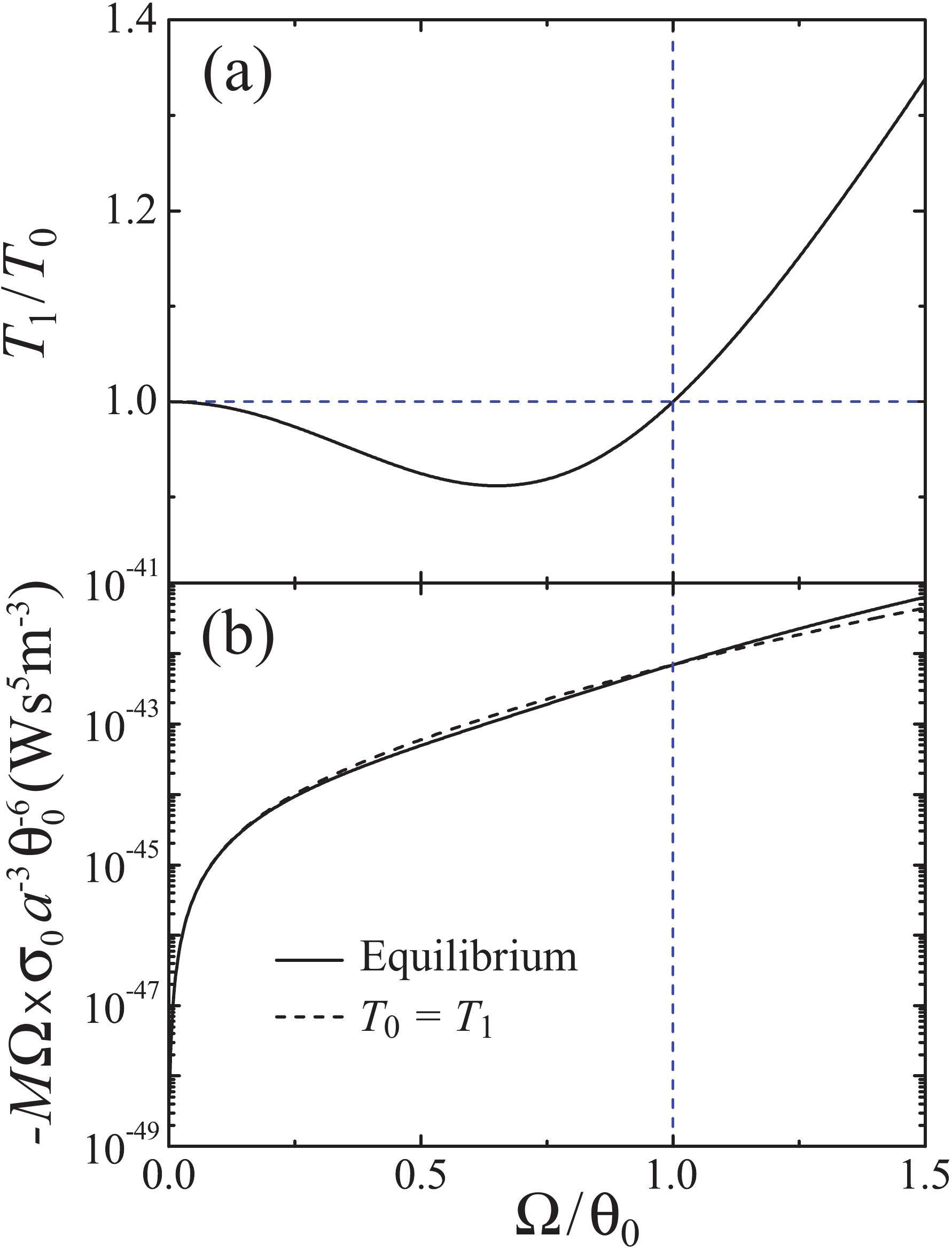}}
\caption{Equilibrium temperature and stopping of a metallic sphere. {\bf (a)} Normalized particle temperature at equilibrium ($T_1/T_0$) as a function of $\Omega/\theta_0$, where $\theta_0=2\pi k_BT_0/\hbar$ (see Fig.\ \ref{Fig1}). {\bf (b)} Universal normalized stopping power both at equilibrium temperatures (solid curve) and at equal temperatures ($T_0=T_1$, broken curve).} \label{Fig2}
\end{figure}

The power absorbed by the particle in the form of thermal heating $P^{\rm abs}$ can be obtained from energy conservation, expressed by the identity $-M\Omega=P^{\rm rad}+P^{\rm abs}$, where the left-hand side represents mechanical energy dissipation (stopping power). Using Eqs.\ (\ref{Pdrude}) and (\ref{Mdrude}), we find
\begin{eqnarray}
P^{\rm abs}_{\rm D}=\frac{\hbar a^3}{120\pi^2c^3\sigma_0}\Big[2\Omega^6+\Omega^2(\theta_0^4-3\theta_1^4)
+\frac{5}{7}(\theta_0^6-\theta_1^6)\Big].
\label{PaDrude}
\end{eqnarray}
The particle equilibrium temperature is determined by the condition $P^{\rm abs}=0$, and it is stable because $\partial P^{\rm abs}/\partial T_1<0$ [this inequality is obvious from Eq.\ (\ref{PaDrude}), but it can be easily derived in the general case from Eqs.\ (\ref{P})-(\ref{M})]. Unlike conventional friction of a spinning object immersed in a fluid, vacuum friction is not always leading to particle heating, as shown in Fig.\ \ref{Fig2}(a) from the solution of $P^{\rm abs}_{\rm D}=0$. Actually, $T_1<T_0$ for finite temperatures and rotation velocities below $\Omega=\theta_0$, whereas particle heating occurs at higher $\Omega$. The crossing point between these two types of behavior is independent of particle size $a$ and conductivity $\sigma_0$.

At $T_0=0$, we find $\theta_1\approx0.867\,\Omega$, so that the $\Omega^5$ dependence of $M_{\rm D}$ is maintained with the particle at equilibrium temperature. The loss of mechanical energy is then fully converted into a radiated power $P^{\rm rad}_{\rm D}\approx0.013\,\hbar a^3\Omega^6/c^3\sigma_0$.

It should be noted that having the particle at equilibrium temperature or at the same temperature as the vacuum results in significant differences in the stopping power [Fig.\ \ref{Fig2}(b), calculated from Eqs.\ (\ref{drude})-(\ref{PaDrude})].

\section{Emission spectra}

\begin{figure}
\centerline{\includegraphics*[width=8cm]{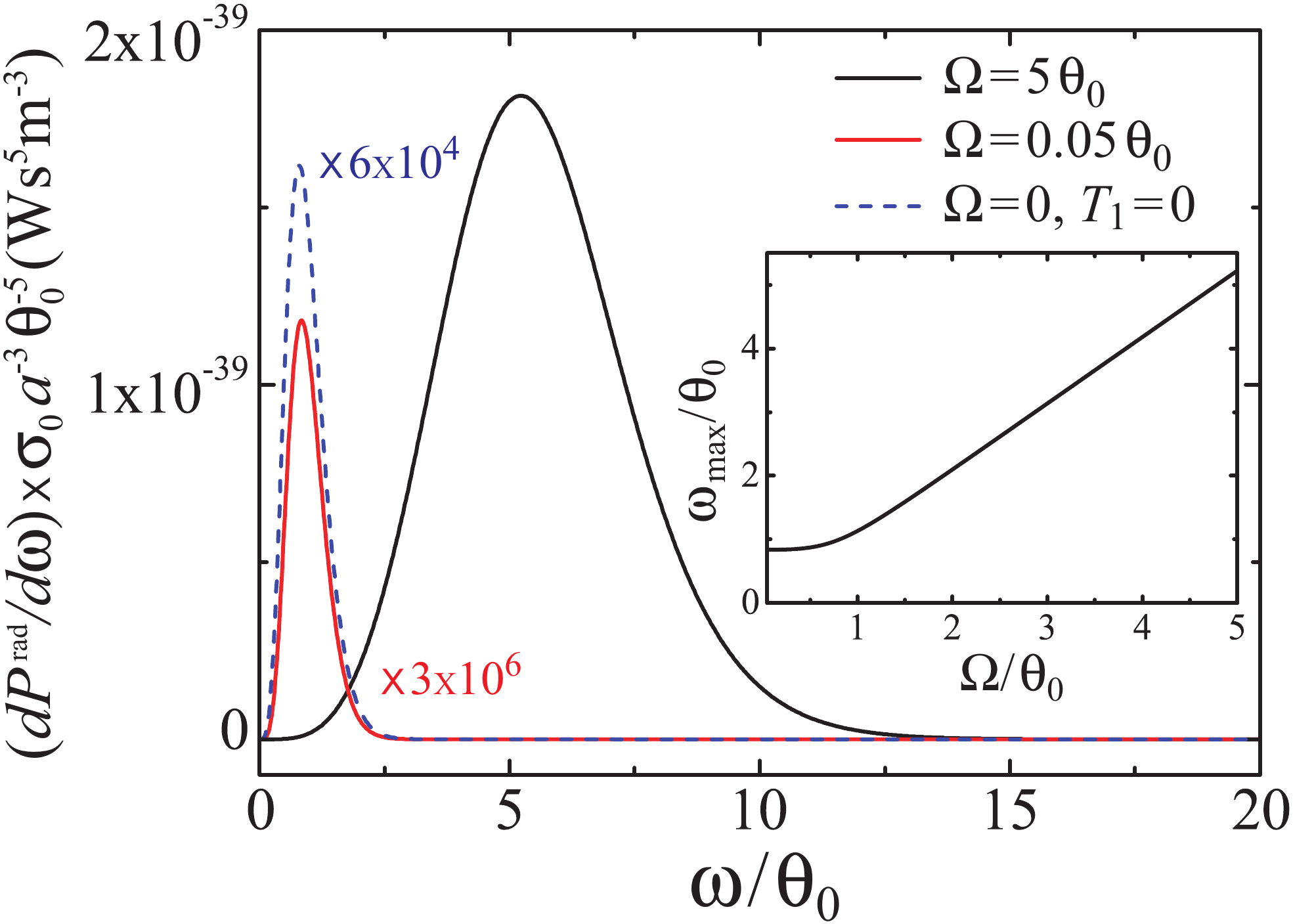}}
\caption{Power spectrum $dP^{\rm rad}/d\omega=\hbar\omega[\Gamma(\omega)-\Gamma(-\omega)]$ [see Eq.\ (\ref{Pw})] radiated by a metallic spinning particle for various rotation frequencies. Solid curves: emission at equilibrium temperatures. Dashed curve: absorption by a particle at rest and $T_1=0$. The emitted-photon frequency $\omega$ is normalized to $\theta_0=2\pi k_BT_0/\hbar$. The inset shows the frequency of maximum emission at equilibrium as a function of $\Omega/\theta_0$.} \label{Fig3}
\end{figure}

The probability of emitting photons at frequency $\omega$ is given by $\Gamma(\omega)-\Gamma(-\omega)$ [see Eq.\ (\ref{Pw})], which is normalized per unit of emission-frequency range. The emission profile at low rotation velocities ($\Omega=0.05\,\theta_0$ curve in Fig.\ \ref{Fig3}) mimics the absorption spectrum from a static cold particle (dashed curve), also peaked around $\hbar\omega\approx5k_BT_1$ for Drude spheres. However, the maximum of emission is driven by $\Omega$ for faster rotations (see inset and $\Omega=5\,\theta_0$ curve in Fig.\ \ref{Fig3}), thus signalling a significant departure from standard black-body theory.

\section{Stopping time}

At low rotation velocity and finite temperature, the frictional torque acting on a metallic particle is proportional to $\Omega$ [see Eq.\ (\ref{Mdrude})]. The correction to the particle equilibrium temperature [$\theta_1\approx\theta_0-(7/15)\Omega^2/\theta_0$] can be then neglected to first order in $\Omega$, so the torque becomes $M\approx-\beta\Omega$, where $\beta=\hbar a^3\theta_0^4/30\pi^2c^3\sigma_0$. From Newton's second law, we find an $\Omega(t)=\Omega(0)\exp(-t/\tau)$ time dependence of the rotation velocity, where $\tau=I/\beta$ is the characteristic stopping time and $I$ is the moment of inertia. For a spherical Drude particle, we find
\begin{eqnarray}
\tau=\frac{(\hbar c)^3}{\pi}\frac{\rho a^2\sigma_0}{(k_BT_0)^4},\label{tau}
\end{eqnarray}
where $\rho$ is the particle density.

\begin{figure}
\centerline{\includegraphics*[width=8cm]{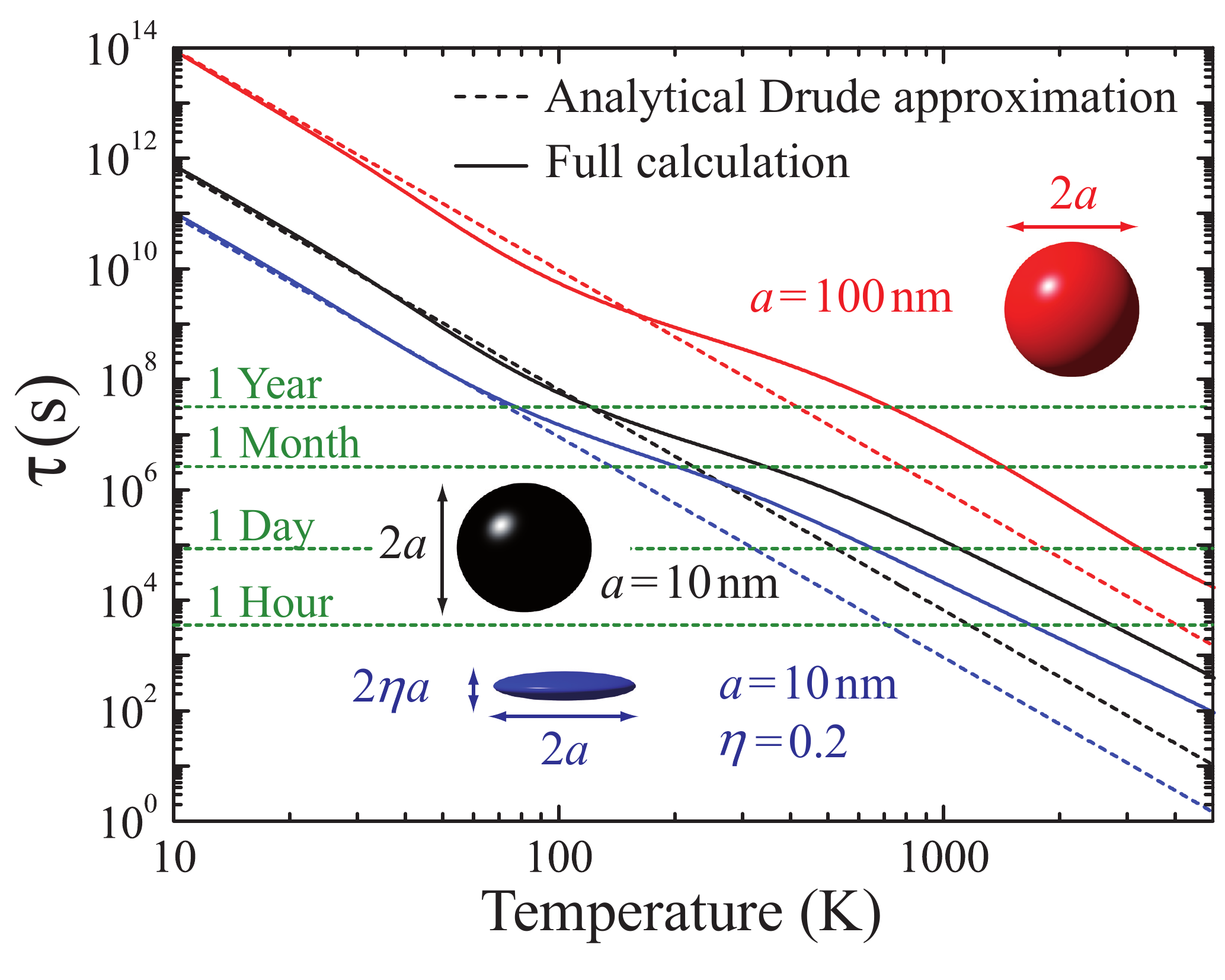}}
\caption{Characteristic stopping time of spinning graphite particles as a function of vacuum temperature. Solid curves: full calculation using measured dielectric functions for the graphite particles.\cite{D03} Broken curves: analytical Drude approximation [Eq. (\ref{tau})]. Various particle sizes and shapes are considered: spheres of radius 10\,nm and 100\,nm, and an oblate ellipsoid of radius 10\,nm and aspect ratio $\eta=0.2$. Low rotation velocities $\Omega\ll k_BT_0/\hbar$ are assumed (e.g., $\Omega\ll21\,$GHz at $T_0=1\,$K).} \label{Fig4}
\end{figure}

Graphite particles are abundant in interstellar dust,\cite{HW1962} so we focus on them as an important case to study the rotation stopping time. The frequency-dependent dielectric function of graphite is taken from optical data,\cite{D03} tabulated for different particle sizes, which differ due to nonlocal corrections. The low-$\omega$ behavior is well approximated by the Drude model with $\sigma_0=2.3\times10^4\;(2.0\times10^5)\,\Omega^{-1}$m$^{-1}$ for spherical particles of radius $a=10\,(100)\,$nm, where the response has been averaged over different crystal orientations. Plugging this into Eq.\ (\ref{tau}), we obtain the results shown in Fig.\ \ref{Fig4} by broken lines. Interband transitions become important in the response of graphite at frequencies above $\hbar\omega\sim10^{-2}\,$eV, so we expect a deviation from Drude behavior at temperatures above $\sim100\,$K in this material. This is indeed confirmed by numerically integrating Eq.\ (\ref{M}) with the full tabulated response of graphite to obtain $\tau$ (Fig.\ \ref{Fig4}, solid curves). For the particle sizes under consideration, stopping times are small on cosmic scales within the plotted range of temperatures, which are often encountered in hot dust regions.\cite{HW1962} In cooler areas ($T_0=2.7\,$K), 100\,nm graphite particles have a stopping time $\tau\sim\,0.6$ billion years.

Dust particles can adopt non-spherical shapes. In particular, for oblate ellipsoids Eq.\ (\ref{drude}) [${\rm Im}\{\alpha(\omega)\}$] must be corrected by a factor $\eta/9L^2$, where $\eta$ is the aspect ratio (see inset in Fig.\ \ref{Fig4}) and $L$ is the depolarization factor for equatorial polarization, approximately linear in $\eta$.\cite{friction5} Also, $I$ is linear in $\eta$, thus leading to a $\tau\propto\eta^2$ dependence for fixed radius. We show in Fig.\ \ref{Fig4} the case $a=10\,$nm and $\eta=0.2$, which exhibits a significant reduction in $\tau$ compared to spherical particles of the same radius. In a related context, translational motion leads to thermal drag,\cite{drag} only at nonzero temperature and with similar stopping times.

\section{Concluding remarks}

The present results can be relevant to study the distribution of rotation velocities of cosmic nanoparticles, which could be eventually examined through measurements of rotational frequency shifts.\cite{BB97MHS05} Besides, relatively small stopping times are predicted for graphite nanoparticles, which ask for experimental corroboration (for example, using in-vacuo optical trapping setups). By analogy to the Purcell effect,\cite{P1946} the frictional torque can be altered due to the presence of physical boundaries that modify the density of states appearing in Eq.\ (\ref{Pw}), thus opening new possibilities for controlling the degree of friction (e.g., the torque can be strongly reduced at low temperature and small rotation frequency by placing the particle inside a metallic cavity, which produces a threshold of $\rho^0$ in $\omega$).

\section*{ACKNOWLEDGMENT}

This work has been supported by the Spanish MICINN (MAT2007-66050 and Consolider NanoLight.es). A.M. acknowledges an FPU scholarship from ME.

\appendix

\section{The fluctuation-dissipation theorem}

The fluctuation-dissipation theorem (FDT) is a well-established result, first formulated by Nyquist\cite{N1928} and later proved by Callen and Welton.\cite{CW1951} It connects the fluctuations of the product of two operators with the dissipation expressed through the imaginary part of their response function. We give a simple derivation here, with a notation that is appropriate to deal with fluctuations of vacuum fields and particle polarizations.

Let us consider a Hamiltonian perturbed by a term
\begin{eqnarray}
\Hh'(t)=-\varphi(t)\qh(t),\nonumber
\end{eqnarray}
where $\varphi(t)$ is a time-dependent function, $\qh(t)$ is an operator in the Heisenberg picture, related to its Schr\"odinger representation $\qh_S$ through $\qh(t)=\exp(i\Hh_0t/\hbar)\qh_S\exp(-i\Hh_0t/\hbar)$, and $\Hh_0$ is the unperturbed Hamiltonian. In the Heisenberg representation, Schr\"odinger's equation becomes $\Hh'|\phi\rangle=i\hbar\partial|\phi\rangle/\partial t$, and we have $|\phi\rangle=\exp(i\Hh_0t/\hbar)|\phi_S\rangle$.

Under the condition $\Hh'(t)_{\overrightarrow{t\rightarrow-\infty}}0$, the eigenstates of the total Hamiltonian become
\begin{eqnarray}
|\phi_m(t)\rangle&=&|m\rangle-\frac{i}{\hbar}\int_{-\infty}^tdt'H'(t')|\phi_m(t')\rangle\nonumber\\
&\approx&|m\rangle-\frac{i}{\hbar}\int_{-\infty}^tdt'H'(t')|m\rangle,\nonumber
\end{eqnarray}
where the last line corresponds to first-order perturbation theory, and $|m\rangle$ is a state of the unperturbed Hamiltonian with energy $E_m$ (i.e., $\Hh_0|m\rangle=E_m|m\rangle$).

\widetext

The expected value of another operator $\ph(t)$ is simply given by
\begin{eqnarray}
\left\langle\ph(t)\right\rangle&=&\frac{1}{Z}\sum_me^{-E_m/k_BT}\left\langle\phi_m(t)|\ph(t)|\phi_m(t)\right\rangle\nonumber\\
&\approx&\frac{1}{Z}\sum_me^{-E_m/k_BT}\left[\left\langle m|\ph(t)|m\right\rangle+\frac{i}{\hbar}\int_{-\infty}^tdt'\varphi(t')\left\langle m|[\ph(t),\qh(t')]|m\right\rangle\right], \label{eq1}
\end{eqnarray}
where
\begin{eqnarray}
Z=\sum_me^{-E_m/k_BT}\nonumber
\end{eqnarray}
is the partition function at temperature $T$. The first term in Eq.\ (\ref{eq1}) reduces to $\langle m|\ph(t)|m\rangle=\langle m|\ph_S|m\rangle$, and from here, we can recast (\ref{eq1}) as
\begin{eqnarray}
\left\langle\delta\ph(t)\right\rangle\equiv\left\langle\ph(t)-\ph(-\infty)\right\rangle=\int dt'\chi(t-t')\varphi(t'), \nonumber
\end{eqnarray}
where
\begin{eqnarray}
\chi(t-t')=\frac{i}{\hbar}\theta(t-t')\frac{1}{Z}\sum_me^{-E_m/k_BT}\left\langle m|[\ph(t),\qh(t')]|m\right\rangle \label{eq2}
\end{eqnarray}
is a susceptibility function. Now, using the closure relation $|n\rangle\langle n|=I$, we can write
\begin{eqnarray}
\left\langle m|[\ph(t),\qh(t')]|m\right\rangle=\sum_n\left[\left\langle m|\ph_S|n\right\rangle\left\langle n|\qh_S|m\right\rangle e^{i(E_m-E_n)(t-t')/\hbar}-\left\langle m|\qh_S|n\right\rangle\left\langle n|\ph_S|m\right\rangle e^{-i(E_m-E_n)(t-t')/\hbar}\right]. \nonumber
\end{eqnarray}
Inserting this back into Eq.\ (\ref{eq2}), taking the time Fourier transform to work in frequency space, and using the identity
\begin{eqnarray}
\int_0^\infty dt\;e^{i\Delta t}=\frac{i}{\Delta+i0^+}, \nonumber
\end{eqnarray}
we find
\begin{eqnarray}
\chi(\omega)&=&\int dt\;\chi(t)\;e^{i\omega t} \nonumber\\
&=&\frac{-1}{Z}\sum_{m,n}\left\langle m|\ph_S|n\right\rangle\left\langle n|\qh_S|m\right\rangle\frac{e^{-E_m/k_BT}-e^{-E_n/k_BT}}{\hbar\omega+E_m-E_n+i0^+}. \nonumber
\end{eqnarray}
Incidentally, the zero-temperature susceptibility reads
\begin{eqnarray}
\chi(\omega)\;\;_{\overrightarrow{T\rightarrow0}}\;\;\;-\sum_m\left[\frac{\left\langle 0|\ph_S|m\right\rangle\left\langle m|\qh_S|0\right\rangle}{\hbar\omega+E_0-E_m+i0^+}
-\frac{\left\langle 0|\qh_S|m\right\rangle\left\langle m|\ph_S|0\right\rangle}{\hbar\omega+E_m-E_0+i0^+}\right]. \nonumber
\end{eqnarray}
Finally, the dissipation associated to $\chi$ can be written
\begin{eqnarray}
{\rm Im}\left\{\chi(\omega)\right\}=\left(1-e^{-\hbar\omega/k_BT}\right)\frac{\pi}{Z}\sum_{m,n}e^{-E_m/k_BT}\left\langle m|\ph_S|n\right\rangle\left\langle n|\qh_S|m\right\rangle\;\delta(\hbar\omega+E_m-E_n). \label{eqdis}
\end{eqnarray}

Similarly, we can write the average over fluctuations as
\begin{eqnarray}
S(t-t')\equiv\left\langle\ph(t)\qh(t')\right\rangle=\frac{1}{Z}\sum_{m,n}e^{-E_m/k_BT}\;e^{i(E_m-E_n)(t-t')/\hbar}\left\langle m|\ph_S|n\right\rangle\left\langle n|\qh_S|m\right\rangle
=\int\frac{d\omega}{2\pi} S(\omega) e^{-i\omega(t-t')}, \nonumber
\end{eqnarray}
where
\begin{eqnarray}
S(\omega)=\frac{2\pi\hbar}{Z}\sum_{m,n}e^{-E_m/k_BT}\;\left\langle m|\ph_S|n\right\rangle\left\langle n|\qh_S|m\right\rangle\;\delta(\hbar\omega+E_m-E_n). \label{eqflu}
\end{eqnarray}

The relation between $S(\omega)$ and ${\rm Im}\left\{\chi(\omega)\right\}$ that one obtains by comparing Eqs.\ (\ref{eqdis}) and (\ref{eqflu}) constitutes the general form of the fluctuation-dissipation theorem:
\begin{eqnarray}
S(\omega)=2\hbar\;[n(\omega)+1]\;{\rm Im}\left\{\chi(\omega)\right\}, \label{Sw}
\end{eqnarray}
where \[n(\omega)=\frac{1}{e^{\hbar\omega/k_BT}-1}\] is the Bose-Einstein distribution function.

We can formulate a more useful relation by noticing that $\left\langle\ph(t)\qh(t')\right\rangle$ is a function of $t-t'$, so that its double Fourier transform satisfies
\begin{eqnarray}
\left\langle\ph(\omega)\qh(\omega')\right\rangle=\int\;dt\,dt'\;e^{i\omega t+i\omega't'} S(t-t')=\int\;d\tau\;e^{i\omega\tau}\;S(\tau)\;\int dt'\;e^{i(\omega+\omega')t'}=2\pi\;\delta(\omega+\omega')\;S(\omega), \nonumber
\end{eqnarray}
and from here we find the expression
\begin{eqnarray}
\left\langle\ph(\omega)\qh(\omega')\right\rangle&=&4\pi\hbar\;[n(\omega)+1]\;{\rm Im}\left\{\chi(\omega)\right\}\;\delta(\omega+\omega'). \label{fdt1}
\end{eqnarray}
Proceeding like above, the Fourier transform of the fluctuation $\left\langle\qh(t')\ph(t)\right\rangle$ reads $\left\langle\qh(\omega')\ph(\omega)\right\rangle=\exp(-\hbar\omega/k_BT)\,S(\omega)$, which together with (\ref{Sw}) leads to
\begin{eqnarray}
\left\langle\qh(\omega')\ph(\omega)\right\rangle&=&4\pi\hbar\;n(\omega)\;{\rm Im}\left\{\chi(\omega)\right\}\;\delta(\omega+\omega'). \label{fdt2}
\end{eqnarray}
Finally, it should be noticed that $\ph(\omega)\qh(\omega')$ is not an observable in general, but the symmetrized product is Hermitian, and therefore, an observable. From Eqs.\ (\ref{fdt1}) and (\ref{fdt2}), we find
\begin{eqnarray}
\frac{1}{2}\left\langle\ph(\omega)\qh(\omega')+\qh(\omega')\ph(\omega)\right\rangle=4\pi\hbar\left[n(\omega)+\frac{1}{2}\right]{\rm Im}\left\{\chi(\omega)\right\}\delta(\omega+\omega').\label{fdt3}
\end{eqnarray}
Equations\ (\ref{fdt1})-(\ref{fdt3}) are general forms of the FDT. Next, we formulate specific applications for dipole and electric-field fluctuations.

\subsection{FDT for fluctuations of the dipole moment}

Now, we can apply the above general expressions of the FDT to dipole-dipole fluctuations, with the identifications
\begin{eqnarray}
\ph(t)&\rightarrow&p_i(t), \nonumber\\
\qh(t)&\rightarrow&p_j(t), \nonumber\\
\chi(t)&\rightarrow&\alpha_{ij}(t), \nonumber\\
\varphi(t)&\rightarrow&E_j(t), \nonumber
\end{eqnarray}
where $p_i$ and $p_j$ are components of the dipole moment along directions $i$ and $j$, respectively, $E_j$ is the electric field along $j$ at the position of the dipole, and $\alpha_{ij}$ is the $(i,j)$ component of the polarizability tensor. The interaction Hamiltonian is $\Hh'(t)=-E_j(t)p_j(t)$, where $E_j$ is regarded as a time-dependent function and $q_j$ as an operator. The susceptibility acts in frequency space according to $\left\langle\delta p_i(\omega)\right\rangle=\alpha_{ij}(\omega)E_j(\omega)$. With these substitutions, the FDT [Eqs.\ (\ref{fdt1})-(fdt3)] takes the forms
\begin{eqnarray}
\left\langle p_i(\omega)p_j(\omega')\right\rangle&=&4\pi\hbar\;[n(\omega)+1]\;{\rm Im}\left\{\alpha_{ij}(\omega)\right\}\;\delta(\omega+\omega'), \nonumber\\
\left\langle p_j(\omega')p_i(\omega)\right\rangle&=&4\pi\hbar\;n(\omega)\;{\rm Im}\left\{\alpha_{ij}(\omega)\right\}\;\delta(\omega+\omega'), \nonumber\\
\frac{1}{2}\left\langle p_i(\omega)p_j(\omega')+p_j(\omega')p_i(\omega)\right\rangle&=&4\pi\hbar\;\left[n(\omega)+\frac{1}{2}\right]\;{\rm Im}\left\{\alpha_{ij}(\omega)\right\}\;\delta(\omega+\omega'). \label{FDTp}
\end{eqnarray}

\subsection{FDT for fluctuations of the electric field}

Similarly, the fluctuations of the electric field can be analyzed with the substitutions
\begin{eqnarray}
\ph(t)&\rightarrow&E_i(\rb,t), \nonumber\\
\qh(t)&\rightarrow&E_j(\rb',t), \nonumber\\
\chi(t)&\rightarrow&G_{ij}(\rb,\rb',t), \nonumber\\
\varphi(t)&\rightarrow&p_j(t), \nonumber
\end{eqnarray}
where $E_i(\rb,t)$ and $E_j(\rb',t)$ are components of the electric field along directions $i$ and $j$ at positions $\rb$ and $\rb'$, respectively. The Green tensor of the electromagnetic field in vacuum is given, in frequency space $\omega$, by
\begin{eqnarray}
G_{ij}(\rb,\rb',\omega)=\frac{\exp(ikR)}{R^3}\,\left[(k^2R^2+ikR-1)\,\delta_{ij}-(k^2R^2+3ikR-3)\,\frac{R_iR_j}{R^2}\right], \label{Gij}
\end{eqnarray}
where $\Rb=\rb-\rb'$ and $k=\omega/c$. The interaction Hamiltonian is again $\Hh'=-p_jE_j$, but now $p_j$ is a time-dependent function and $E_j$ is an operator. The susceptibility acts as
\begin{eqnarray}
\left\langle\delta E_i(\omega)\right\rangle=G_{ij}(\rb,\rb',\omega)\,p_j(\omega), \nonumber
\end{eqnarray}
which is consistent with the definition of Eq.\ (\ref{Gij}).\cite{J99} The FDT takes the forms
\begin{eqnarray}
\left\langle E_i(\rb,\omega)E_j(\rb',\omega')\right\rangle&=&4\pi\hbar\;[n(\omega)+1]\;{\rm Im}\left\{G_{ij}(\rb,\rb',\omega\right\}\;\delta(\omega+\omega'), \nonumber\\
\left\langle E_j(\rb',\omega')E_i(\rb,\omega)\right\rangle&=&4\pi\hbar\;n(\omega)\;{\rm Im}\left\{G_{ij}(\rb,\rb',\omega)\right\}\;\delta(\omega+\omega'), \nonumber\\
\frac{1}{2}\left\langle E_i(\rb,\omega)E_j(\rb',\omega')+E_j(\rb',\omega')E_i(\rb,\omega)\right\rangle&=&4\pi\hbar\;\left[n(\omega)+\frac{1}{2}\right]\;{\rm Im}\left\{G_{ij}(\rb,\rb',\omega)\right\}\;\delta(\omega+\omega').\nonumber
\end{eqnarray}
Interestingly, for $\rb=\rb'$ we have ${\rm Im}\left\{G_{ij}(\rb,\rb,\omega)\right\}=(2\pi^2\omega/3)\rho^0\delta_{ij}$, where $\rho^0$ is the photonic local density of states ($\rho^0=\omega^2/\pi^2c^3$ in vacuum), which yields
\begin{eqnarray}
\frac{1}{2}\left\langle E_i(\rb,\omega)E_j(\rb,\omega')+E_j(\rb,\omega')E_i(\rb,\omega)\right\rangle&=&\frac{8\pi^3\hbar\omega\rho^0}{3}\;\left[n(\omega)+\frac{1}{2}\right]\;\delta_{ij}\;\delta(\omega+\omega').\label{FDTE}
\end{eqnarray}

\section{Polarizability of a rotating particle}

Rotational motion affects the polarizability $\alpha$ of a small spinning particle. This quantity represents bubble diagrams made up of virtual (de-)excitations. The corresponding matrix elements involve particle ground and excited states of electronic and vibrational nature, which do rotate with the particle. Therefore, $\alpha$ provides the relation between fields and dipoles expressed in the rotating frame. The external electric field has to be transformed to the rotating frame before multiplying by $\alpha$ to obtain the induced dipole, and this in turn has to be transformed back to the rest frame.

We assume rotation around the $z$ axis, and denote by $\xx$ and $\yy$ the unit vectors in the lab frame, and $\xx'$ and $\yy'$ the unit vectors in the rotating frame. With the particle rotating with angular frequency $\Omega$, the angle between $\xx'$ and $\xx$ increases with time as $\Omega t$. Then, expressing the external electric field in frequency space $\omega$, we can write
\begin{eqnarray}
\Eb(t)&=&\int\frac{d\omega}{2\pi}e^{-i\omega t}[E_x(\omega)\xx+E_y(\omega)\yy]\nonumber\\
&=&\int\frac{d\omega}{2\pi}e^{-i\omega t}[E'_x(\omega)\xx'+E'_y(\omega)\yy'].\nonumber
\end{eqnarray}
The relation between the field in the lab and rotating frames can be obtained by writing $\xx$ and $\yy$ in terms of $\xx'$ and $\yy'$, and then absorbing the $\Omega t$ dependence in the time exponentials within the above integral. We obtain
\begin{eqnarray}
E'_x(\omega)&=&\frac{1}{2}\left[E_x(\omega_+)+E_x(\omega_-)-iE_y(\omega_+)+iE_y(\omega_-)\right],\nonumber\\
E'_y(\omega)&=&\frac{1}{2}\left[iE_x(\omega_+)-iE_x(\omega_-)+E_y(\omega_+)+E_y(\omega_-)\right],\nonumber
\end{eqnarray}
where $\omega_\pm=\omega\pm\Omega$. Besides, the induced dipole in the rotating frame is $p'_x(\omega)=\alpha_\perp(\omega)E'_x(\omega)$ and $p'_y(\omega)=\alpha_\perp(\omega)E'_y(\omega)$, where $\alpha_\perp$ is the static-particle polarizability for polarization perpendicular to the rotation axis. We assume the particle to have axial symmetry around $z$, so that polarization along that direction, described by $\alpha_\parallel$, is unaffected by the rotational motion. Finally, we need to transform the induced dipole from the rotating frame to the lab frame. Proceeding like above, we have
\begin{eqnarray}
p_x(\omega)&=&\frac{1}{2}\left[p'_x(\omega_+)+p'_x(\omega_-)+ip'_y(\omega_+)-ip'_y(\omega_-)\right],\nonumber\\
p_y(\omega)&=&\frac{1}{2}\left[-ip'_x(\omega_+)+ip'_x(\omega_-)+p'_y(\omega_+)+p'_y(\omega_-)\right].\nonumber
\end{eqnarray}
Putting all this together, we find the induced dipole in the rotating frame to have the same frequency as the external field for this particular case of axially-symmetric particles. More precisely, the relation between electric field and induced dipole in the lab frame is described by an effective polarizability $\alpha^{\rm eff}(\omega)$ according to
\begin{eqnarray}
\begin{bmatrix}p_x(\omega)\\p_y(\omega)\\p_z(\omega)\end{bmatrix}=
\begin{bmatrix}
\alpha_{xx}^{\rm eff}(\omega) & \alpha_{xy}^{\rm eff}(\omega) & 0 \\
\alpha_{yx}^{\rm eff}(\omega) & \alpha_{yy}^{\rm eff}(\omega) & 0 \\
0 & 0 & \alpha_{zz}^{\rm eff}(\omega)
\end{bmatrix}
\cdot
\begin{bmatrix}E_x(\omega)\\E_y(\omega)\\E_z(\omega)\end{bmatrix},
\nonumber
\end{eqnarray}
where
\begin{subequations}\label{alphaeff}
\begin{eqnarray}
\alpha^{\rm eff}_{xx}(\omega)&=&\alpha^{\rm eff}_{yy}(\omega)=\frac{1}{2}[\alpha_\perp(\omega+\Omega)+\alpha_\perp(\omega-\Omega)],\label{alphaeffa}\\
\alpha^{\rm eff}_{xy}(\omega)&=&-\alpha^{\rm eff}_{yx}(\omega)=\frac{i}{2}[\alpha_\perp(\omega+\Omega)-\alpha_\perp(\omega-\Omega)],\label{alphaeffb}\\
\alpha^{\rm eff}_{zz}(\omega)&=&\alpha_\parallel(\omega). \label{alphaeffc}
\end{eqnarray}
\end{subequations}
Like $\alpha$, this effective polarizability satisfies the retarded response condition $\alpha^{\rm eff}(-\omega)=\left[\alpha^{\rm eff}(\omega)\right]^*$.

\section{Derivation of the vacuum torque}

The torque exerted by an electric field $\Eb$ on a dipole $\pb$ is given by
\begin{eqnarray}
{\bf M}=\pb\times\Eb.\label{MM0}
\end{eqnarray}
This is intuitively understood by considering the dipole to be formed by a negative point charge located at the origin and a positive point charge on which the electric field produces a force; the position vector of the positive charge enters both the dipole and the torque, thus yielding Eq.\ (\ref{MM0}). The vacuum torque can be obtained from (\ref{MM0}), with $\pb$ and $\Eb$ produced by polarization and field fluctuations, respectively. Noticing that these two kinds of fluctuations are uncorrelated, the vacuum torque for a particle with symmetry of revolution around its rotation axis $z$ is given by
\begin{eqnarray}
M&=&\zz\cdot\left\langle\pb^{\rm fl}\times\Eb^{\rm ind}+\pb^{\rm ind}\times\Eb^{\rm fl}\right\rangle\nonumber\\
&=&\left\langle p_x^{\rm fl}E_y^{\rm ind}-p_y^{\rm fl}E_x^{\rm ind}+p_x^{\rm ind}E_y^{\rm fl}-p_y^{\rm ind}E_x^{\rm fl}\right\rangle,\label{MM}
\end{eqnarray}
where the average is taken over fluctuations of the vacuum electric field $\Eb^{\rm fl}$ and the particle polarizability $\pb^{\rm fl}$.

With the particle at the origin, we obtain the induced field in frequency space from $\Eb^{\rm ind}(\rb,\omega)=G(0,\rb,\omega)\cdot\pb^{\rm fl}(\omega)$, where the components of $G$ are given by Eq. (\ref{Gij}) and the $r\rightarrow0$ limit is to be understood in what follows. Similarly, the induced dipole in (\ref{MM}) is obtained from $\pb^{\rm ind}(\omega)=\alpha^{\rm eff}(\omega)\cdot\Eb^{\rm fl}(0,\omega)$. Writing the fields and the dipoles in frequency space and using these equations for the induced fields and dipoles, we obtain from (\ref{MM})
\begin{eqnarray}
M=\int\frac{d\omega d\omega'}{(2\pi)^2}\,e^{-i(\omega+\omega')t}&&\bigg[
G_{yx}(0,\rb,\omega)\left\langle p_x^{\rm fl}(\omega)p_x^{\rm fl}(\omega')\right\rangle+G_{yy}(0,\rb,\omega)\left\langle p_y^{\rm fl}(\omega)p_x^{\rm fl}(\omega')\right\rangle\nonumber\\
&&-G_{yy}(0,\rb,\omega)\left\langle p_y^{\rm fl}(\omega)p_y^{\rm fl}(\omega')\right\rangle-G_{xy}(0,\rb,\omega)\left\langle p_x^{\rm fl}(\omega)p_y^{\rm fl}(\omega')\right\rangle\nonumber\\
&&+\alpha_{xy}^{\rm eff}(\omega)\left\langle E_y^{\rm fl}(0,\omega)E_y^{\rm fl}(0,\omega')\right\rangle+\alpha_{xx}^{\rm eff}(\omega)\left\langle E_x^{\rm fl}(0,\omega)E_y^{\rm fl}(0,\omega')\right\rangle\nonumber\\
&&-\alpha_{yx}^{\rm eff}(\omega)\left\langle E_x^{\rm fl}(0,\omega)E_x^{\rm fl}(0,\omega')\right\rangle-\alpha_{yy}^{\rm eff}(\omega)\left\langle E_x^{\rm fl}(0,\omega)E_y^{\rm fl}(0,\omega')\right\rangle\bigg].\nonumber
\end{eqnarray}
Explicit expressions for $G$ and $\alpha^{\rm eff}$ are taken from Eqs.\ (\ref{Gij}) and (\ref{alphaeff}), respectively. Finally, we apply the FDT from Eqs.\ (\ref{FDTp}) and (\ref{FDTE}), assuming that the products of the $p$ and $E$ observables are symmetrized as explained at the end of the FDT section. The field fluctuations occur in the vacuum at a temperature $T_0$, so that the corresponding FDT yields a Bose-Einstein distribution at that temperature, which we denote $n_0$. Likewise, the polarization fluctuations take place in the particle at temperature $T_1$, and thus one obtains a distribution $n_1$. After some straightforward algebra, we find that only the imaginary part of $G$ survives in the above integral. Finally, the torque reduces to
\begin{eqnarray}
M=-\int_{-\infty}^\infty d\omega\;\hbar\Gamma(\omega),\label{Mint}
\end{eqnarray}
where
\begin{eqnarray}
\Gamma(\omega)&=&(2\pi\omega\rho^0/3)\,2{\rm Im}\{\alpha_\perp(\omega-\Omega)\}\,\big[n_1(\omega-\Omega)-n_0(\omega)\big]\label{gam}
\end{eqnarray}
is a spectral distribution function (see below). A similar result is obtained for particles without axial symmetry, but with orthogonal axes of polarization, and rotating around one of them, $z$; in this case, after time averaging, we find that $\alpha_\perp$ has to be substituted by the average of the polarizability over directions perpendicular to the rotation axis, $(\alpha_{xx}+\alpha_{yy})/2$.

A similar derivation can be carried out for the torque produced by magnetic polarization, which is negligible for very small particles in which the magnetic polarizability is generally small, although it can play a role for metallic particles of large conductivity, leading to imaginary parts of the electric and magnetic polarizabilities that are comparable in magnitude (the magnetic effect can be even dominant).

\section{Derivation of the radiated power}

Following the intuitive explanation of the torque acting on a dipole (see previous section), we can argue that the work exerted on a dipole by an electric field is $P=\Eb\cdot\partial\pb/\partial t$. Here, we are interested in the radiation produced by friction acting on the particle, or equivalently, the work done by the particle on the vacuum, which we can write in terms of fluctuating and induced dipoles and fields as
\begin{eqnarray}
P^{\rm rad}=-\left\langle\Eb^{\rm ind}\cdot\partial\pb^{\rm fl}/\partial t+\Eb^{\rm fl}\cdot\partial\pb^{\rm ind}/\partial t\right\rangle.\nonumber
\end{eqnarray}
After lengthy but straightforward algebra, mimicking the steps followed in the calculation of the vacuum torque, we obtain
\begin{eqnarray}
P^{\rm rad}=\int_{-\infty}^\infty\hbar\omega\;d\omega\;\Gamma(\omega)\nonumber,
\end{eqnarray}
where $\Gamma(\omega)$ is the same as in Eq.\ (\ref{gam}). Now, we can interpret $\Gamma$ as the spectral distribution of the rate of emission (when $\omega\Gamma>0$) or absorption ($\omega\Gamma<0$). It should be noted that $\omega$ in these expressions has been carefully ensured to correspond to the photon frequency in the lab frame.

Incidentally, we are missing in $\Gamma$ the contribution of polarization along the rotation axis $z$, which can be easily derived from half the contribution of radial polarization for a non-rotating particle. The full result becomes
\begin{eqnarray}
\Gamma(\omega)=(2\pi\omega\rho^0/3)\,\Big\{2\,{\rm Im}\{\alpha_\perp(\omega-\Omega)\}\,\big[n_1(\omega-\Omega)-n_0(\omega)\big]+{\rm Im}\{\alpha_\parallel(\omega)\}\,\big[n_1(\omega)-n_0(\omega)\big]\Big\}.\label{gamtrue}
\end{eqnarray}
Actually, the new term inside the curly brackets, together with the leading prefactor, yield an odd function of $\omega$, so that this term vanishes under the integral of Eq.\ (\ref{Mint}), and therefore, we can use Eq.\ (\ref{gamtrue}) both for the torque and for the radiated power.

\section{Quantum-mechanical coupling of rotational motion and free photons}

In a quantum-mechanical description of rotational vacuum friction, the interaction between the particle and the free-space electromagnetic field can be described by the Hamiltonian\cite{L1983}
\begin{eqnarray}
\Hh'=i\sum_j \sqrt{\frac{2\pi\hbar\omega_j}{V}}(\db\cdot\eh_j)\;[\ah_j-\ah_j^+],\label{Hj}
\end{eqnarray}
where the sum runs over photon states $j$ of polarization vectors $\eh_j$ and frequencies $\omega_j$, $V$ is the quantization volume, $\ah_j^+$ and $\ah_j$ are photon creation and annihilation operators, respectively, and $\db$ is the particle-dipole operator. We have considered the particle to be small compared to the wavelength of the involved photons, so that the use of the dipole approximation is justified.

The rotational state of a spinning particle can be described as a combination of eigenfunctions $\exp(im\varphi)/\sqrt{2\pi}$, where $\varphi$ is the rotation angle and $m$ is the azimuthal quantum number. For rotation velocity $\Omega$, the values of $m$ are peaked around $m\sim I\Omega/\hbar$, where $I$ is the moment of inertia. The angle $\varphi$ enters Eq.\ (\ref{Hj}) through the transformation of the dipole operator from the lab frame ($\db$) to the rotating frame ($\db'$). More precisely,
\begin{eqnarray}
d_x&=&d'_x\cos\varphi-d'_y\sin\varphi,\nonumber\\
d_y&=&d'_x\sin\varphi+d'_y\cos\varphi.\nonumber
\end{eqnarray}
The terms in $\sin\varphi$ and $\cos\varphi$ can produce transitions $m\rightarrow m\pm1$, while $d'_x$ and $d'_y$ generate internal transitions in the particle. This is accompanied by the emission or absorption of one photon, according to Eq.\ (\ref{Hj}). In other words, rotational friction is associated to transitions involving simultaneous changes in (i) the rotational state of the particle, (ii) its internal state, and (iii) the number of photon states. All of these elements must be present in the interaction described by Eq.\ (\ref{Hj}). This implies that non-absorbing particles (at least within the relevant range of photon frequencies) cannot undergo friction, because they do not possess the excited states that are necessary to sustain internal excitations.


\begin{thebibliography}{27}
\bibitem[{fri({\natexlab{c}})}]{friction2}
\bibinfo{note}{From a quantum mechanical point of view, the lowest-order
  coupling Hamiltonian producing vacuum friction includes terms with
  simultaneous changes in (i) the rotational state of the particle
  ($m\rightarrow m\pm 1$ jumps in the azimuthal quantum number), (ii) the
  internal state of the particle, and (iii) the number of vacuum photon states (see Appendix).
  In non-absorbing spheres, no such matrix elements exist
  because there are not internal excitations available.}
\end{thebibliography}

\end{document}